\documentclass{article}
\usepackage{spconf,amsmath,epsfig}
\usepackage{amssymb,color}
\usepackage{times,amsmath,epsfig,textcomp,mathrsfs}

\begin{document}
\ninept
\title{Low-Complexity Robust Data-Adaptive Dimensionality Reduction
Based on Joint Iterative Optimization of Parameters \vspace{-0.5em}}

\name{Peng Li \dag ~and Rodrigo C. de Lamare*$^\#$ \vspace{-0.8em}}

\address{\normalsize \dag Communications Research Lab., TU-Ilmenau, Germany, 98684\\
\normalsize *Department of Electronics, The University of York, England, YO10 5BB\\
{\normalsize $^\#$CETUC, Pontifical Catholic University of Rio de Janeiro, Brazil}\\
\normalsize Emails: peng.li@tu-ilmenau.de, {
delamare@cetuc.puc-rio.br} \vspace{-1em} \sthanks{This work was
supported by UK MOD under contract from the Centre for Defence
Enterprise}}

\maketitle
\thispagestyle{empty}
\pagestyle{empty}

% \IEEEpeerreviewmaketitle

\begin{abstract}

%In dynamic propagation environments with large sensor arrays,
%beamforming algorithms may suffer from strong interference, steering
%vector mismatches, a low convergence speed and a high computational
%complexity. Reduced-rank signal processing techniques provide a way
%to address the problems mentioned above.
This paper presents a low-complexity robust data-dependent
dimensionality reduction based on a modified joint iterative
optimization (MJIO) algorithm for reduced-rank beamforming and
steering vector estimation. The proposed robust optimization
procedure jointly adjusts the parameters of a rank-reduction matrix
and an adaptive beamformer. The optimized rank-reduction matrix
projects the received signal vector onto a subspace with lower
dimension. The beamformer/steering vector optimization is then
performed in a reduced-dimension subspace. We devise efficient
stochastic gradient and recursive least-squares algorithms for
implementing the proposed robust MJIO design. The proposed robust
MJIO beamforming algorithms result in a faster convergence speed and
an improved performance. Simulation results show that the proposed
MJIO algorithms outperform some existing full-rank and reduced-rank
algorithms with a comparable complexity.

\end{abstract}

\section{Introduction}

Adaptive beamforming algorithms often encounter problems when they
operate in dynamic environments with large sensor arrays. These
include snapshot deficiency, steering vector mismatches caused by
calibration and pointing errors, and a high computational
complexity. In terms of complexity, an expensive inverse operation
of the covariance matrix of the received data is often required,
resulting in a high computational complexity that may prevent the
use of adaptive beamforming in important applications like sonar and
radar. In order to overcome this computational complexity issue,
adaptive versions of the linearly constrained beamforming algorithms
such as minimum variance distortionless response (MVDR) with
stochastic gradient and recursive least squares \cite{BK:H2002} have
been extensively reported. These adaptive algorithms estimate the
data covariance matrix iteratively and the complexity is reduced by
recursively computing the weights. However, in a dynamic environment
with large sensor arrays such as those found in radar and sonar
applications, adaptive beamformers with a large number of array
elements may fail in tracking signals embedded in strong
interference and noise. The convergence speed and tracking
properties of adaptive beamformers depend on the size of the sensor
array and the eigen-spread of the received covariance matrix
\cite{BK:H2002}. Regarding the steering vector mismatches often
found in practical applications of beamforming, they are responsible
for a significant performance degradation of algorithms. Prior work
on robust beamforming design \cite{LSW03,VGL03,Sam11} has considered
different strategies to mitigate the effects of these mismatches. An
effective method to deal with mismatches is the Robust Capon
Beamforming (RCB) technique of \cite{LSW03}. A key limitation of
\cite{LSW03} and other robust techniques \cite{VGL03,Sam11,locsme}
is their high cost for large sensor arrays and their suitability to
dynamic environments.

Reduced-rank signal processing techniques \cite{Sam11}-\cite{SLPL13}
provide a way to address some of the problems mentioned above.
Reduced-dimension methods are often needed to speed-up the
convergence of beamforming algorithms and reduce their computational
complexity. They are particularly useful in scenarios in which the
interference lies in a low-rank subspace and the number of degrees
of freedom required the mitigate the interference through
beamforming is significantly lower than that available in the sensor
array. In reduced-rank schemes, a rank-reduction matrix is
introduced to project the original full-dimension received signal
onto a lower dimension. {The advantage of reduced-rank methods lie
in their superior convergence and tracking performance achieved by
exploiting the low-rank nature of the signals.} It offers a large
reduction in the required number of training samples over full-rank
methods \cite{BK:H2002}. Several reduced-rank strategies for
processing data collected from a large number of sensors have been
reported in the last few years, which include beamspace methods
\cite{Sam11}, Krylov subspace techniques \cite{Wang,SLPL13}, and
methods based on joint and iterative optimization of parameters
\cite{delamare_jidf,LWF10,FLW10}.

{Despite the improved convergence and tracking performance achieved
with Krylov methods \cite{Wang,SLPL13}, they are relatively complex
and may suffer from numerical problems. On the other hand, the joint
optimization technique reported in \cite{LWF10} outperforms the
Krylov-based method with efficient adaptive implementations.
However, this algorithm suffers from the problem of rank one. In
order to address this problem,} in this paper, we introduce a
low-complexity robust data-dependent dimensionality reduction based
on a modified joint iterative optimization (MJIO) algorithm for
reduced-rank beamforming and steering vector estimation. The
proposed MJIO design strategy jointly optimizes the rank-reduction
matrix and a reduced-rank beamformer, which ensures that the
rank-reduction matrix has a desired rank. {Another contribution of
this work is the introduction of a bank of perturbed steering
vectors as candidate array steering vectors around the true steering
vector. The candidate steering vectors are responsible for
performing rank reduction and the reduced-rank beamformer forms the
beam in the direction of the signal of interest (SoI). We devise
efficient stochastic gradient(SG) and recursive least-squares (RLS)
algorithms for implementing the proposed robust MJIO design.
Simulation results show that the proposed MJIO algorithms outperform
existing full-rank and reduced-rank algorithms with a comparable
complexity.

This paper is organized as follows. The system model is described in
Section 2. The reduced-rank MVDR beamforming with MJIO is formulated
in Section 3. A robust version of MJIO is investigated in Section 4
and simulations are discussed in Section 5.

\section{System Model and Problem Statement}

Let us consider a uniform linear array (ULA) with $M$ sensor
elements, which receive $K$ narrowband signals where $K \leq M$. The
DoAs of the $K$ signals are {$\theta_0, \ldots \theta_{K-1}$}. The
received vector $\boldsymbol{x}[i] \in \mathbb{C}^{M \times 1}$ at
the $i$-th snapshot (time instant), can be modelled as
\begin{equation}
 \boldsymbol{x}[i] = \boldsymbol{A}(\boldsymbol{\theta})\boldsymbol{s}[i]+\boldsymbol{n}[i], \qquad i = 1, \ldots, N
\end{equation}
where $\boldsymbol{\theta} = [\theta_0, \ldots, \theta_{K-1}]^T \in \mathbb{R}^{K \times 1}$ convey the DoAs of the $K$ signal sources. $\boldsymbol{A}(\boldsymbol{\theta}) = [\boldsymbol{a}(\theta_0),\ldots,\boldsymbol{a}(\theta_{K-1})] \in \mathbb{C}^{M \times K}$ comprises $K$ steering vectors which are given as
\begin{equation}
 \boldsymbol{a}(\theta_k) = [1, e^{-2\pi j \frac{\iota}{\lambda_c}\cos(\theta_k)},\ldots,e^{-2\pi j (M-1) \frac{\iota}{\lambda_c}\cos(\theta_k)}]^T.
\end{equation}
where $\lambda_c$ is the wavelength and $\iota$ is the inter-element
distance of the ULA. The $K$ steering vectors
$\boldsymbol{a}\{\boldsymbol{\theta_k}\} \in \mathbb{C}^{M \times
1}$ are assumed to be linearly independent. The source data are
modelled as $\boldsymbol{s} \in \mathbb{C}^{K\times1}$ and
$\boldsymbol{n}[i]\in \mathbb{C}^{M \times 1}$ is the noise vector,
which is assumed to be zero-mean, $N$ is assumed to be the
observation size and $[i]$ denotes the time instant. For full-rank
processing, the adaptive beamformer output for the SoI is written as
\begin{equation}
 y_k[i] = \boldsymbol{\omega}_k^H[i]\boldsymbol{x}[i],
\end{equation}
where the beamformer $\boldsymbol{\omega}_k \in
\mathbb{C}^{M\times1}$ is derived according to a design criterion.
The optimal weight vector is obtained by maximizing the
signal-to-interference-plus-noise ratio (SINR) and
\begin{equation}
\mbox{SINR}_{\scriptsize \mbox{opt}} = \frac{\boldsymbol{\omega}_{\scriptsize \mbox{opt}}^H\boldsymbol{R}_k\boldsymbol{\omega}_{\scriptsize \mbox{opt}}}{\boldsymbol{\omega}_{\scriptsize \mbox{opt}}^H\boldsymbol{R}_{i+n}\boldsymbol{\omega}_{\scriptsize \mbox{opt}}},
\end{equation}
where $\boldsymbol{R}_k$ and $ \boldsymbol{R}_{i+n}$ denote the SoI
and  interference-plus-noise covariance matrices, respectively. {
Full-rank beamformers usually suffer from high complexity and low
convergence speed. In the following, we focus on the design of
low-complexity reduced-dimension beamforming algorithms.}

\section{Dimension Reduction with Modified JIO}

In this section we describe reduced-rank algorithms based on the
proposed MJIO design of beamformers. The scheme jointly optimizes a
rank-reduction matrix and a reduced-rank beamformer that operates at
the output of the projection matrix. The bank of adaptive
beamformers in the front-end is responsible for performing
dimensionality reduction, which is followed by a reduced-rank
beamformer which effectively forms the beam in the direction of the
SoI. This two-stage scheme allows the adaptation with different
update rates, which could lead to a significant reduction in the
computational complexity per update. Specifically, this complexity
reduction can be obtained as the dimensionality reduction performed
by the rank-reduction matrix could be updated less frequently than
the reduced-rank beamformer. The design criterion of the proposed
MVDR-MJIO beamformer is given by the optimization problem
\begin{equation}
\begin{aligned}
\min_{\boldsymbol{\boldsymbol{\omega},\boldsymbol{s}_d}} \qquad &
\boldsymbol{\omega}^H\boldsymbol{S}_D^H\boldsymbol{R}\boldsymbol{S}_D\boldsymbol{\omega}, \\
 \mbox{subject to} \qquad &  \boldsymbol{\omega}^H\sum_{d=1}^{D}
 \boldsymbol{q}_d \boldsymbol{s}_d^H \boldsymbol{a}_d = 1,
 \label{co_mjio}
\end{aligned}
\end{equation}
where $\boldsymbol{R}$ is the covariance matrix obtained from
sensors, vector $\boldsymbol{q}_d$ with dimension $D \times 1$ is a
zero vector except its $d$-th element been one. The vector
$\boldsymbol{s}_d \in \mathbb{C}^{M \times 1} $ is the $d$-th column
of the projection matrix $\boldsymbol{S}_D \in \mathbb{C}^{M \times
D}$. The vectors $\boldsymbol{a}_d, d = 1 \ldots D$ represent the
assumed steering vector and $D-1$ small perturbations of the assumed
steering vector. {Each recursion updates a different column of
$\boldsymbol{S}_D$. An increased rank of $\boldsymbol{S}_D$ is
required for higher $d$, and the rank one problem in \cite{LWF10}
can be avoided. The constrained optimization problem in
(\ref{co_mjio}) can be solved by using the method of Lagrange
multipliers \cite{LSW03}.} The Lagrangian of the MVDR-MJIO design is
expressed by
\begin{equation}\label{eq:mvdr-mjio}
 f(\boldsymbol{\omega},\boldsymbol{s}_d) = E\Big{\{}\Big{|}\boldsymbol{\omega}^H\sum_{d=1}^{D} \boldsymbol{q}_d \boldsymbol{s}_d^H \boldsymbol{x}\Big{|}^2\Big{\}} + \lambda \Big{(} \boldsymbol{\omega}^H\sum_{d=1}^{D} \boldsymbol{q}_d \boldsymbol{s}_d^H \boldsymbol{a} - 1 \Big{)}.
\end{equation}

\subsection{\bf{Stochastic Gradient Adaptation}}

In this subsection, we present a low-complexity SG \cite{BK:H2002}
adaptive reduced-rank algorithm for efficient implementation of the
MJIO algorithm. By computing the instantaneous gradient terms of
(\ref{eq:mvdr-mjio}) with respect to $\boldsymbol{\omega}[i]^*$ and
$\boldsymbol{s}_d[i]^*$, we obtain
\begin{equation}
 \boldsymbol{\omega}[i+1] =  \boldsymbol{\omega}[i] - \mu_w \boldsymbol{P}_w[i]\boldsymbol{S}_D^H[i] \boldsymbol{x}[i] z^*[i],
\end{equation}
\begin{equation}
 \boldsymbol{s}_d[i+1] =  \boldsymbol{s}_d[i] - \mu_s \boldsymbol{P}_s[i]\boldsymbol{x}[i] z^*[i] w_d^*[i], d = 1, \ldots, D,
\end{equation}
where $w_d$ is the $d$th element of the reduced-rank beamformer
${\boldsymbol \omega}[i]$ and the projection matrices that enforce
the constraints are
\begin{equation}
\boldsymbol{P}_w[i] = \boldsymbol{I}_D -
(\boldsymbol{a}_D^H[i]\boldsymbol{a}_D[i])^{-1}\boldsymbol{a}_D[i]\boldsymbol{a}_D^H[i],
\end{equation}
and
\begin{equation}
\boldsymbol{P}_s[i] = \boldsymbol{I}_M -
(\boldsymbol{a}^H[i]\boldsymbol{a}[i])^{-1}\boldsymbol{a}[i]\boldsymbol{a}^H[i],
\end{equation}
the scalar $z^*[i] =
\boldsymbol{x}^H[i]\boldsymbol{S}_D[i]\boldsymbol{\omega}[i]=\tilde{\boldsymbol
x}^H[i]{\boldsymbol \omega}$ and
\begin{equation}
\boldsymbol{a}_D[i] = \sum_{d=1}^{D} \boldsymbol{q}_d \boldsymbol{s}_d[i]^H \boldsymbol{a}[i] \in \mathbb{C}^{D \times 1}.
\end{equation}
is the estimated steering vector in reduced dimension. The
calculation of $\boldsymbol{P}_\omega[i]$ requires a number of $D^2+
D+1$ complex multiplications, the computation of
$\boldsymbol{P}_s[i]$ and $z[i]$ requires $D^2+DM+M+1$ and $DM+D$
complex multiplications, respectively. Therefore, we can conclude
that for each iteration, the SG adaptation requires
$4MD+4D^2+3D+M+6$ complex multiplications.

\subsection{\bf{Recursive Least Squares Adaptation}}

Here we derive an adaptive reduced-rank RLS {\cite{BK:H2002}} type
algorithm for efficient implementation of the MVDR-MJIO method. The
reduced-rank beamformer $\boldsymbol{\omega}[i]$ is updated as
follows:
\begin{equation}
\boldsymbol{\omega}[i] = \frac{\boldsymbol{R}^{-1}_D[i]
\boldsymbol{a}_D[i]}{\boldsymbol{a}_D^H[i] \boldsymbol{R}^{-1}_D[i]
\boldsymbol{a}_D[i]}, \label{omega_rec}
\end{equation}
where
\begin{equation}
\tilde{\boldsymbol{k}}[i+1] =
\frac{\alpha^{-1}\boldsymbol{R}_D^{-1}[i]\tilde{\boldsymbol{x}}[i+1]}{1+\alpha^{-1}\tilde{\boldsymbol{x}}^H[i+1]\boldsymbol{R}_D^{-1}[i]\tilde{\boldsymbol{x}}[i]},
\end{equation}
\begin{equation}
 {\boldsymbol{R}}_D^{-1}[i+1] = \alpha^{-1}{\boldsymbol{R}}_D^{-1}[i] - \alpha^{-1}
\tilde{\boldsymbol{k}}[i+1] \tilde{\boldsymbol{x}}^H[i+1]
{\boldsymbol{R}}_D^{-1}[i],
\end{equation}
The columns $\boldsymbol{s}_d[i]$ of the rank-reduction matrix  are
updated by
\begin{equation}
\boldsymbol{s}_d[i] =
\frac{\boldsymbol{R}^{-1}[i]\boldsymbol{a}_d[i]\boldsymbol{a}_d^H[i]
\boldsymbol{\beta}_d[i]} {\boldsymbol{a}_d^H[i]
\boldsymbol{R}^{-1}[i]\boldsymbol{a}_d[i]w_d[i]}, d = 1, \ldots, D,
\end{equation}
where $\boldsymbol{\beta}_d[i] = \sum_{d=1}^{D}\boldsymbol{s}_d[i] w_d[i] - \sum_{l=1, l\neq d}^{D}\boldsymbol{s}_l[i] w_l[i]$ and
\begin{equation}
\boldsymbol{k}[i+1] = \frac{\alpha^{-1}\boldsymbol{R}^{-1}[i]\boldsymbol{x}[i+1]}{1+\alpha^{-1}\boldsymbol{x}^H[i+1]\boldsymbol{R}^{-1}[i]\boldsymbol{x}[i]},
\end{equation}
\begin{equation}\label{eq:riccati02}
\boldsymbol{R}^{-1}[i+1] = \alpha^{-1} \boldsymbol{R}^{-1}[i] - \alpha^{-1} \boldsymbol{k}[i+1]\boldsymbol{x}^H[i+1]\boldsymbol{R}^{-1}[i],
\end{equation}
where $0 \ll \alpha < 1$ is the forgetting factor. The inverse of
the covariance matrix $\boldsymbol{R}^{-1}$ is obtained recursively.
Equation (\ref{eq:riccati02}) is initialized by using an identity
matrix $\boldsymbol{R}^{-1}[0] = \delta \boldsymbol{I}$ where
$\delta$ is a positive constant. The computational complexity of the
proposed adaptive reduced-rank RLS type MVDR-MJIO method requires
$4M^2 + 3D^2 + 3D + 2$ complex multiplications. The MVDR-MJIO
algorithm has a complexity significantly lower than a full-rank
scheme if a low rank ($D \ll M$) is selected.

\section{Proposed Robust Capon MJIO Beamforming}

In this section, we present a robust beamforming method based on the
Robust Capon Beamforming (RCB) technique reported in \cite{LSW03}
and the MJIO detailed in the previous section for robust beamforming
applications with large sensor arrays. The proposed technique,
denoted Robust Capon Beamforming MJIO (RCB-MJIO), gathers the
robustness of the RCB approach \cite{LSW03} against uncertainties
and the low-complexity of MJIO techniques. Assuming that the DoA
mismatch is within a spherical uncertainty set, the proposed
RCB-MJIO technique solves the following optimization problem:
\begin{equation}
\begin{aligned}
   \min_{\boldsymbol{\boldsymbol{a}_d,\boldsymbol{s}_d}} \qquad &  \boldsymbol{a}_d^H\boldsymbol{S}_D^H\boldsymbol{R}^{-1}\boldsymbol{S}_D\boldsymbol{a}_d, \\
   \mbox{subject to} \qquad &  \left\|\boldsymbol{S}_D^H\boldsymbol{a}_d - \boldsymbol{S}_D^H\bar{\boldsymbol{a}}\right\|^2 = \epsilon, \\
\end{aligned}
\end{equation}
where $\bar{\boldsymbol{a}}$ is the assumed steering vector and
$\boldsymbol{a}_d$ is the updated steering vector for each
iteration. The constant $\epsilon$ is related to the radius of the
uncertainty sphere. The Lagrangian of the RCB-MJIO constrained
optimization problem is expressed by
\begin{equation}\label{eq:rcb-mjio}
\begin{aligned}
 f_{\tiny \mbox{RCB}}(\boldsymbol{a}_d,\boldsymbol{s}_d) & = \left(\sum_{d =1}^D \boldsymbol{q}_d \boldsymbol{s}_d^H\boldsymbol{a}_d\right)^H\boldsymbol{R}_D^{-1}\left(\sum_{d =1}^D \boldsymbol{q}_d \boldsymbol{s}_d^H\boldsymbol{a}_d\right) + \\
 &\lambda_{\tiny \mbox{RCB}} \left( \left\|  \sum_{d =1}^D \boldsymbol{q}_d \boldsymbol{s}_d^H\boldsymbol{a}_d - \sum_{d =1}^D \boldsymbol{q}_d \boldsymbol{s}_d^H\bar{\boldsymbol{a}}       \right\|^2 - \epsilon \right),
\end{aligned}
\end{equation}
where $\boldsymbol{R}_D^{-1} = \boldsymbol{S}_D^H
\boldsymbol{R}^{-1}\boldsymbol{S}_D $ is the reduced rank covariance
matrix. From the above Lagrangian, we will devise efficient adaptive
beamforming algorithms in what follows.

\subsection{Stochastic Gradient Adaptation}

We devise an SG adaptation strategy based on the alternating
minimization of the Lagrangian in (\ref{eq:rcb-mjio}), which yields
\begin{equation}
\begin{aligned}
 \tilde{\boldsymbol{a}}_d[i+1] = \tilde{\boldsymbol{a}}_d [i] - \mu_a[i] \boldsymbol{g}_a[i],\\
\boldsymbol{s}_d[i+1] = \boldsymbol{s}_d [i] - \mu_s[i]
\boldsymbol{g}_s[i], \label{rec_a&g}
\end{aligned}
\end{equation}
where $\mu_a[i]$ and $\mu_s[i]$ are the step-sizes of the SG
algorithms, the parameter vectors $\boldsymbol{g}_a[i]$ and
$\boldsymbol{g}_s[i]$ are the partial derivatives of the Lagrangian
in (\ref{eq:rcb-mjio}) with respect to
$\tilde{\boldsymbol{a}}_d^*[i]$ and $\boldsymbol{s}_d^*[i]$,
respectively. The recursion for $\boldsymbol{g}_a[i]$ is given by
\begin{equation}
 \boldsymbol{g}_a[i] = \left( \frac{1}{\lambda}_{\tiny \mbox{RCB}[i]} \boldsymbol{S}_D^H[i] \boldsymbol{R}^{-1}[i]\boldsymbol{S}_D[i] + \boldsymbol{I}_D \right) ^{-1} \boldsymbol{S}_D^H[i] \tilde{\boldsymbol{a}}_d[i],
\end{equation}
where
\begin{equation}
\begin{aligned}
 \boldsymbol{g}_s[i] = \boldsymbol{a}_d[i] & \check{\boldsymbol{a}}_d^H[i]  \boldsymbol{r}_d [i] + \tau_d [i] \boldsymbol{a}_d[i]  \boldsymbol{a}_d^H[i] \boldsymbol{s}_d[i],  \\
 & + \lambda_{\tiny \mbox{RCB}}[i]\boldsymbol{\alpha}_d[i]  \boldsymbol{\alpha}_d^H [i] \boldsymbol{s}_d[i],
\end{aligned}
\end{equation}
and
\begin{equation}
\tilde{\boldsymbol{a}}_d = \sum_{d =1}^D \boldsymbol{q}_d \boldsymbol{s}_d^H \boldsymbol{a}_d = \boldsymbol{S}_D^H\boldsymbol{a}_d \in \mathbb{C}^{D \times 1},
\end{equation}
\begin{equation}
\check{\boldsymbol{a}}_d = \sum_{l =1, l\neq d }^D \boldsymbol{q}_l \boldsymbol{s}_l^H \boldsymbol{a}_l \in \mathbb{C}^{D \times 1}.
\end{equation}
We denote $\boldsymbol{\alpha}_d \in \mathbb{C}^{M \times 1}$ as the
difference between the updated steering vectors and the assumed one. The scalar $\tau_d$ is the $d$-th diagonal element of
$\boldsymbol{R}_D^{-1}$. The term $\boldsymbol{r}_d$ denotes the
$d$-th column vector of $\boldsymbol{R}_D^{-1}$. The Lagrange
multiplier obtained is expressed as
\begin{equation}
 \lambda_{\tiny \mbox{RCB}}[i] = -
\left(\boldsymbol{S}_D[i]^H{\boldsymbol{\alpha}}_d[i]
{\boldsymbol{\alpha}}_d^H[i] \boldsymbol{s}_d[i] \right)^{\dag}
{\boldsymbol{R}_D^{-1}[i] \tilde{\boldsymbol{a}}_d[i]
\boldsymbol{a}_d^H[i] \boldsymbol{s}_d[i] }, \label{lamb_sg}
\end{equation}
%\textcolor{red}{Peng, we need to include one or two sentences about
%the computational complexity}
The proposed RCB-MJIO SG algorithm corresponds to (7)-(9) and
(\ref{rec_a&g})-(\ref{lamb_sg}). The calculation of $\lambda_{\tiny
\mbox{RCB}}$ requires $MD + D^2 + 4M + D$ complex multiplications,
and the computation of $\boldsymbol{g}_a[i]$ and
$\boldsymbol{g}_s[i]$ needs $D^3 + MD + D$ and $5M + D + 2$
multiplications, respectively.

\subsection{Recursive Least Squares Adaptation}

We derive an RLS version of the RCB-MJIO method. The steering vector
and the columns of rank-reduction matrix are updated as
\begin{equation}\label{eq:rls1}
\tilde{\boldsymbol{a}}_d[i]  = \left[ \tilde{\boldsymbol{a}}_d[i] - \left( \boldsymbol{I}_D + \lambda_{\tiny \mbox{RCB}}[i] \boldsymbol{R}_D^{-1}[i] \right)^{-1} \tilde{\boldsymbol{a}}_d[i] \right],
\end{equation}
\begin{equation}
 \boldsymbol{s}_d = - \left(\tau_d[i] \boldsymbol{a}_d[i]\boldsymbol{a}_d^H[i] + \lambda_{\tiny \mbox{RCB}}[i] \boldsymbol{\alpha}_d[i]\boldsymbol{\alpha}_d^H[i]\right)^{-1} {\boldsymbol{a}_d[i] \check{\boldsymbol{a}}_d^H[i] \boldsymbol{r}_d[i]},
\end{equation}
\begin{equation}
\tilde{\boldsymbol{k}}[i+1] = \frac{\alpha^{-1}\boldsymbol{R}_D^{-1}[i]\tilde{\boldsymbol{x}}[i+1]}{1+\alpha^{-1}\tilde{\boldsymbol{x}}^H[i+1]\boldsymbol{R}_D^{-1}[i]\tilde{\boldsymbol{x}}[i]},
\end{equation}
\begin{equation}\label{eq:riccati2}
\boldsymbol{R}_D^{-1}[i+1] = \alpha^{-1} \boldsymbol{R}_D^{-1}[i] - \alpha^{-1} \tilde{\boldsymbol{k}}[i+1]\tilde{\boldsymbol{x}}^H[i+1]\boldsymbol{R}_D^{-1}[i],
\end{equation}
where (\ref{eq:rls1})-(\ref{eq:riccati2}) need $2D^3 + 7D^2 + 4D +
3$ complex multiplications, and the projection operations need a
complexity of $MD$ complex multiplications. It is obvious that the
complexity is significantly decreased if the selected rank $D \ll
M$. The proposed RCB-MJIO RLS algorithm employs (\ref{omega_rec})
and (\ref{eq:rls1})-(\ref{eq:riccati2}). The key of the RCB-MJIO RLS
algorithm is to update the assumed steering vector
$\tilde{\boldsymbol{a}}_d[i]$ with RLS iterations, and the updated
beamformer ${\boldsymbol \omega}[i]$ is obtained by plugging
(\ref{eq:rls1}) into (\ref{omega_rec}) without significant extra
complexity.

Note that the complexity introduced by the pseudo-inverse operation
can be removed if $\boldsymbol{S}_D$ has orthogonal column vectors,
this can be achieved by incorporating the Gram-Schmidt procedure in
the calculation of $\boldsymbol{S}_D$. Furthermore, an alternative
recursive realization of the robust adaptive linear constrained
beamforming method introduced by \cite{E2008} can be used to further
reduce the computational complexity requirement to obtain the
diagonal loading terms.

%\textcolor{red}{Peng, can you please check the recursions needed to
%implement the RLS version? How do you compute the recursion for
%${\boldsymbol \omega}[i]$? I gather it is using
%(\ref{omega_rec})...}

\section{Simulations}

In this section, we consider simulations for a ULA with
{$\lambda_c/2$} spacing between the sensor elements and arrays with
$64$ and $320$ sensor elements. The covariance matrix
$\hat{\boldsymbol{R}}$ is obtained by time-averaging recursions with
$N=1,\ldots,120$ snapshots, we use the spherical uncertainty set and
the upper bound is set to $\epsilon = 140$ for $64$ sensor elements
and $\epsilon = 800$ for $320$ sensor elements. There are $4$
incident signals while the first is the SoI, the other $3$ signals'
relative power with respect to the SoI and their DoAs in degrees are
detailed in Table I. The algorithms are trained with $120$ snapshots
and the Signal-to-Noise Ratio (SNR) is set to $10$ dB for all the
simulations.

In Fig.~\ref{fig:SINR_RO}, we select $D = 2$ for rank reduction, the
proposed RCB-MJIO method with the RLS algorithm is used to obtain
the inverse of the covariance matrix $\hat{\boldsymbol{R}}^{-1}[i]$
for each snapshot. We introduce a maximum of $2$ degrees of DOA
mismatch which is independently generated by a uniform random
generator in each simulation run. A non-orthogonal Krylov projection
matrix $\boldsymbol{S_D}[i] \in \mathbb{C}^{ 64 \times 2}$ and a
non-orthogonal MJIO rank-reduction matrix {is} also generated for
rank reduction. {$\boldsymbol{S_D}[i]$ is initialized as
$\boldsymbol{S}_D[0] = [\boldsymbol{I}_D^T,
\boldsymbol{0}_{D\times(M-D)}^T]$.} In Fig.\ref{fig:SINR_RO_nom}, we
choose a similar scenario but without DOA mismatch. We can see from
the plots that the the MJIO and Krylov algorithms have a superior
SINR performance to other existing methods and this is particularly
noticeable for a reduced
number of snapshots. %This is especially noticeable for a reduced
%number of snapshots as they benefit from rank reduction.

In Fig.~\ref{fig:SINR_RO_320} we compare the output SINRs of the
Krylov and the proposed MJIO rank reduction technique using a
spherical constraint in the presence of steering vector errors with
$320$ sensor elements. We assume a DOA mismatch with $2$ degrees and
$4$ interferences with the profile listed in Table I. With Krylov
and MJIO rank-reduction, the MVDR-Krylov, MVDR-MJIO, RCB-Krylov and
RCB-MJIO have superior SINR performance and a faster convergence
compared with their full-rank rivals.

\begin{table}
\footnotesize \centering \caption{\footnotesize Interference and DoA
Scenario, {P(dB) relative to desired user1 / DoA (degree)}}
\label{tab: simulation}
\begin{tabular}{l c c c c c}
\hline \\
\bf {Snapshots} &  \bf{signal1 (SoI)} &  \bf{signal 2}  & \bf{signal 3} &  \bf{signal 4}  \\
\hline   \\
1-120  & 10/90  & 20/35 & 20/135 & 20/165 \\
%1001-2000 & 0/93  & 30/120 & 34/140 & 6/104 & 9/68\\
\hline
\end{tabular}
\end{table}
\vspace{-1.2em}

\begin{figure}[!htb]
\begin{center}
\def\epsfsize#1#2{0.825\columnwidth}
\epsfbox{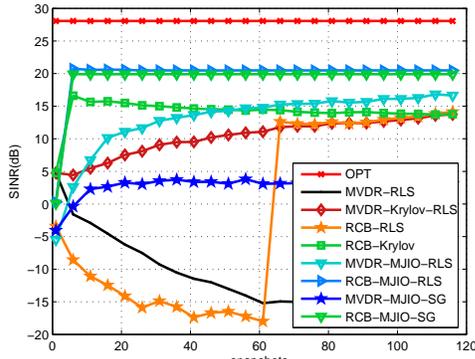} \vspace{-2.2em} \caption{\footnotesize SINR
performance vs. number of snapshots, with steering vector mismatch
due to 2 \textdegree DoA mismatch. Spherical uncertainty set is
assumed for robust beamformers $\epsilon = 140$ ( RLS indicates the
value $\hat{\boldsymbol{R}}^{-1}$ is obtained by using RLS
adaptation), non-orthogonal $\boldsymbol{S_D}[i] \in \mathbb{C}^{ 64
\times 2}$ projection matrix. } \label{fig:SINR_RO}
\end{center}
\end{figure}

\begin{figure}[!htb]
\begin{center}
\def\epsfsize#1#2{0.825\columnwidth}
\epsfbox{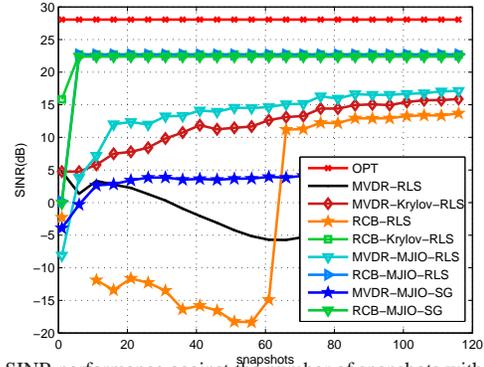} \vspace{-2.2em} \caption{\footnotesize SINR
performance against the number of snapshots without steering vector
mismatch. } \label{fig:SINR_RO_nom}
\end{center}
\end{figure}

\begin{figure}[!htb]
\begin{center}
\def\epsfsize#1#2{0.825\columnwidth}
\epsfbox{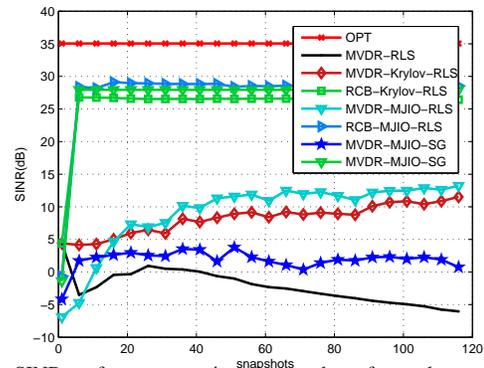} \vspace{-2.2em} \caption{\footnotesize SINR
performance against the number of snapshots with steering vector
mismatch due to 2 \textdegree DoA mismatch. The spherical
uncertainty set is assumed for robust beamformers with $\epsilon =
800$, non-orthogonal $\boldsymbol{S_D}[i] \in \mathbb{C}^{ 320
\times 2}$ rank-reduction matrix. } \label{fig:SINR_RO_320}
\end{center}
\end{figure}

%\newpage

%
\end{document}